\documentclass{svproc}

\usepackage[T1]{fontenc}
\usepackage{graphicx}
\usepackage{subfigure}

\usepackage{stmaryrd}

\usepackage{listings}
\usepackage[english]{babel}
\usepackage{graphicx}
\usepackage{colortbl}
\usepackage{fancyhdr}
\usepackage{listings}
\graphicspath{{figs/}}

\begin{document}
\title{Networks' modulation: How different structural network properties affect the global synchronization of coupled Kuramoto oscillators.}

\author{Juliette Courson\inst{1,2,3}
\and
Thanos Manos\inst{2}
\and
Mathias Quoy\inst{2,4}
}
\authorrunning{Courson et al.}
\institute{Laboratoire de Physique Théorique et Modélisation (LPTM), CNRS, UMR 8089, CY Cergy Paris Université, Cergy-Pontoise Cedex, France \\
\email{juliette.courson@cyu.fr}
\and Equipes Traitement de l’Information et Systèmes (ETIS), CNRS, UMR 8051, ENSEA, CY Cergy Paris Université, Cergy-Pontoise Cedex, France \\
\email{thanos.manos@cyu.fr}
\and Department of Computer Science, University of Warwick, Coventry, UK 
\and IPAL CNRS Singapore\\
\email{mathias.quoy@cyu.fr}}

\maketitle
\begin{abstract}
In a large variety of systems (biological, physical, social etc.), synchronization occurs when different oscillating objects tune their rhythm when they interact with each other. The different underlying network defining the connectivity properties among these objects drives the global dynamics in a complex fashion and affects the global degree of synchrony of the system. Here we study the impact of such types of different network architectures, such as Fully-Connected, Random, Regular ring lattice graph, Small-World and Scale-Free in the global dynamical activity of a system of coupled Kuramoto phase oscillators. We fix the external stimulation parameters and we measure the global degree of synchrony when different fractions of nodes receive stimulus. These nodes are chosen either randomly or based on their respective strong/weak connectivity properties (centrality, shortest path length and clustering coefficient). Our main finding is, that in Scale-Free and Random networks a sophisticated choice of nodes based on their eigenvector centrality and average shortest path length exhibits a systematic trend in achieving higher degree of synchrony. However, this trend does not occur when using the clustering coefficient as a criterion. For the other types of graphs considered, the choice of the stimulated nodes (randomly vs selectively using the aforementioned criteria) does not seem to have a noticeable effect.

\end{abstract}
\section{Introduction}

Complex networks' theory is a powerful tool in various fields that allow us to investigate and understand the real world \cite{Barrat_book_2008,Nicolis_book_2012}. For example, different ensembles of neurons connected by synapses coordinate their activity to perform certain tasks (in biology), infrastructures like the Internet are formed by routers and computer cables and optical fibers (in hardware communication) and the human personal or professional relationships (in social sciences) to name a few \cite{Dorogovtsev_book_2003}.

Nonlinearity is a very important feature in complex systems giving a rich repertoire of different activity patterns, such as stable, unstable, periodic etc. A modification of some parameter might also produce a change in their stability, and therefore in the dynamics of the system. Furthermore, such systems may have a high sensitivity to initial conditions, or to any external input, that could completely change their dynamics \cite{Strogatz_book2015}. 

Such dynamics often yield to a self-organised coherent activity, i.e. to synchronization. The latter can be loosely defined as the capacity of different oscillating objects to adjust their rhythm due to their interaction and plays a key role in a large variety of systems, whether biological, physical, or even social (see e.g. \cite{Pikovsky_book_2001}). In a more formal way, synchronization emerges from the interaction of several autonomous oscillators, also called self-sustained oscillators. That is, nonlinear dynamical systems that produce oscillations without any need of external source. Their dynamics is given by a nonlinear differential equation or, in the case of multiple coupled oscillators, by several coupled differential equations.

The relative way that autonomous oscillators are connected within a given network can affect their global activity and synchronization properties. Neural networks can be represented as a graph of connections between the different neurons. Since the introduction of small-world networks and scale-free networks (see e.g. \cite{WattsStrogatz_1998,BarabasiReka_1999}), the field of network graph analysis has attracted the attention of many studies aimed to better understand complex systems (see e.g. \cite{Jeong_etal_2000,Strogatz2001,LiCai_2004,BassettBullmore_2006,BQ06}). Furthermore, modern network connectivity techniques allow us to capture various aspects of their topological organization, as well as to quantify the local contributions of individual nodes and edges to network's functionality (see e.g. \cite{Sporn_book_2010}).

In neuroscience, synchronization plays a very important role. The human brain is a very large and complex system whose activity comprises the rapid and precise integration of a gigantic amount of signals and stimulus to perform multiple tasks (see e.g. \cite{Sporn_book_2010,Chialvo_2010,Fornito_book_2016}). One example occurs in epileptic seizures, where periods of abnormal synchronization in the neural activity can spread within different regions of the brain, and cause an attack in the affected person (see e.g. \cite{Wong_etal_1986}). More examples are found in other brain diseases such as Parkinson disease, where an excessively synchronized activity in a brain region correlates with motor deficit (see e.g. \cite{Brown_2003,Manos_etal_2021} and references therein) or tinnitus (see e.g. \cite{EggermontTass_2015,Manos_etal_2018a,Manos_etal_2018b} and references therein). %

In this study, we focus at a rather theoretical framework. We set out to investigate the impact of different network architectures, such as Fully-Connected, Random, Regular ring lattice graph, Small-World and Scale-Free in the global dynamical activity of a system of coupled Kuramoto phase oscillators \cite{Kura_1984}. The Kuramoto model has been broadly used to study various types of oscillatory complex activity, see e.g. \cite{Acebron_etal_2005,Rodrigues_etal_20161,Pop_etal_2021} (to name only a few) and references therein. Our goal is to investigate the impact of the network (graph) structure in the system's global degree of synchronization when applying identical and fixed external stimulus to different subsets of nodes which are chosen according to various network connectivity criteria. We find that, in scale-free and random networks, a sophisticated choice of nodes based on graph connectivity properties exhibits a systematic trend in achieving higher degree of synchrony. For the other types of graphs considered, the choice of the stimulated nodes (randomly vs selectively using the aforementioned criteria) seems to not have a noticeable effect.

\section{Methods and Materials}

\subsection{Connectivity measurements} \label{sect:conn}

We here study the dynamics of phase oscillators coupled via binary, undirected graphs $G=(V,E)$, containing a set of $N$ vertices $V = \{v \in \llbracket 1:N \rrbracket\}$ and a set $E = \{(v,w) \in \llbracket 1:N \rrbracket^2\}$ of edges. Let $A$ be the corresponding adjacency matrix, with $A_{vw}=1$ if there is a connection between node $v$ and node $w$, $0$ otherwise. Self-connections are excluded, so $A_{v,v} = 0$ for any vertex $v$. For our analysis later on, we will use the following graph connectivity measurements \cite{Newman_book_2010}:
\begin{itemize}
\item \textbf{Shortest path length.} The shortest path length $L_{v,w}$ between any two nodes $v$ and $w$ is the number of connections on the shortest path going from one to another, computed following Dijkstra's algorithm. We define the shortest path length of a node $v$ as the average shortest path between $v$ and any other node of the network:
\begin{equation} \label{eq:short_path}
    <L_v> = \sum_{w \in V}\frac{ L_{v,w}}{N}.
\end{equation}
Note that $L_{v,w}$ might not be defined if there is no way connecting node $v$ to node $w$. The lower the shortest path length, the fastest the information goes from one node to another. For example, when building a subway network (that is, a graph where different stations are interconnected), one might want to minimize the stations' average shortest path length so the users can easily navigate across the city.

\item \textbf{Centrality.} The eigenvector centrality is used to quantify the importance of a node in the network. Let $\lambda$ be the highest eigenvalue for $A$, so that all the corresponding eigenvector's components are non null. The eigenvector centrality $x_v$ of vertex $v$ is defined as the $v^{th}$ component the eigenvector, namely:
\begin{equation} \label{eq:centra}
    x_v = \frac{1}{\lambda} \sum_{w\in V} A_{w,v} x_w.
\end{equation}
Keeping in mind the subway network example, a station with a high centrality would be densely connected to other stations, in particular to other central ones.
\item \textbf{Clustering.} Let $k_v=\sum_w A_{vw}$ be the degree of node $v$. In the case of a undirected graph, $\frac{k_v(k_v-1)}{2}$
edges can exist in the direct neighborhood of $v$. With $n_v$ the number edges that actually exist in this neighborhood, the local clustering coefficient is defined as \cite{Fornito_book_2016}:
\begin{equation} \label{eq:cluste_coef}
    C_v= \frac{2n_v}{k_v(k_v-1)}.
\end{equation}
That is, in a subway network where stations $B$ and $C$ are the next stop after station $A$ on their line, clustering would give the probability that there exists a line directly connecting $B$ and $C$.
\end{itemize}

\subsection{Neural networks as graphs} \label{sect:graphs}

We investigate synchronization properties in various network configurations that exhibit in general different characteristics. In more detail we here employ the neural networks described in the following list (see e.g. \cite{WattsStrogatz_1998}):

\begin{itemize}
\item \textbf{Fully-Connected networks.} They contains $(N-1)^2$ edges connecting every node in one layer to every node in the other layer.
\item \textbf{Regular networks.} They consist of a lattice of $N$ nodes, each being connected to their $k$ nearest neighbors. 
\item \textbf{Small-World networks.} A Small-World network is constructed from a Regular one after multiple random rewiring phases: going clockwise over the lattice, a vertex and the edge to its nearest neighbor are selected. The edge is removed, and the vertex reconnected to a random node with probability $p$, without duplicating any existing edge. This random rewiring is repeated, considering the following nearest neighbour, until having rewired with probability $p$ all edges of the network. Choosing $p$ in an adequate range of values, the built network exhibits both high mean clustering and low mean characteristic path length, that is small-worldness.
\item \textbf{Random networks.} Using the same procedure as for Small-World networks, setting $p=1$ produces a Random graph, with all edges being systematically randomly rewired. 
\begin{figure}[h]
\centering
    \begin{tabular}{ll}
        \subfigure[Fully-Connected]{\label{sub1} \includegraphics[scale=0.18]{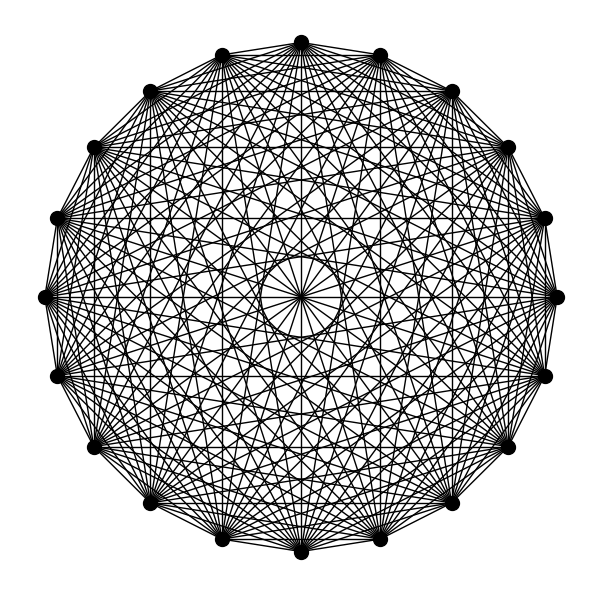}} & \subfigure[Scale-Free]{\label{sub2} \includegraphics[scale=0.18]{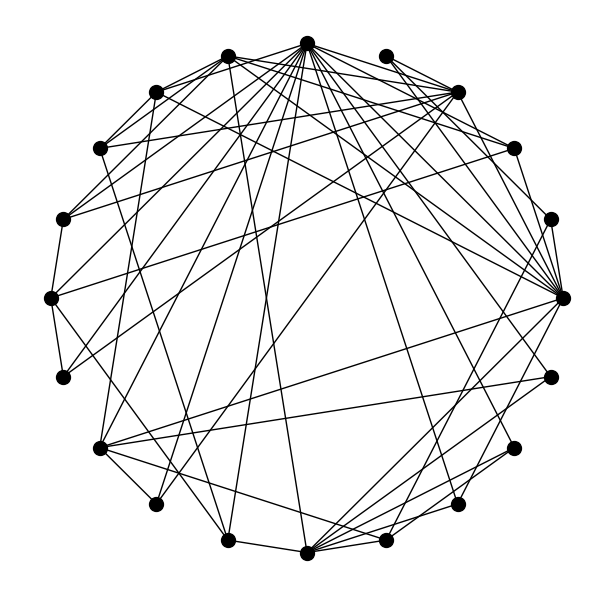}} \\
    \end{tabular}   
    \begin{tabular}{lll}
        \subfigure[Regular]{\label{sub3} \includegraphics[scale=0.18]{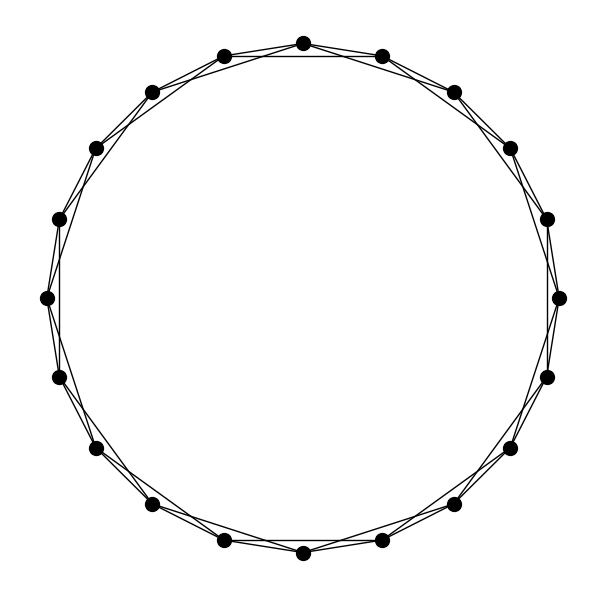}} & \subfigure[Swall-World]{\label{sub4} \includegraphics[scale=0.18]{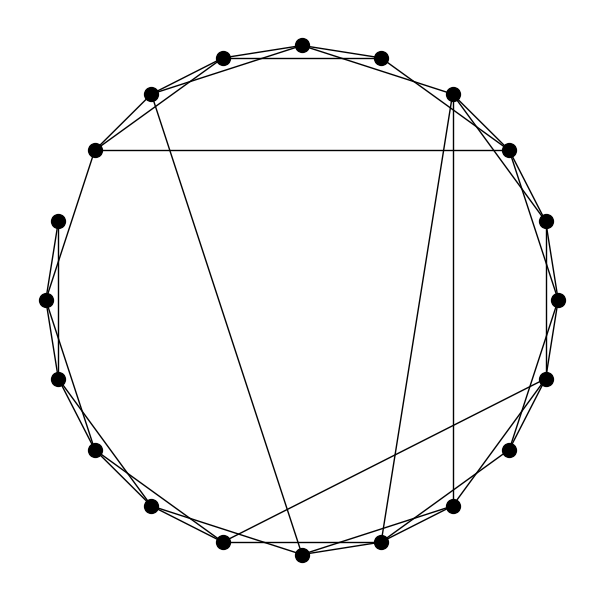}} & \subfigure[Random]{\label{sub5} \includegraphics[scale=0.18]{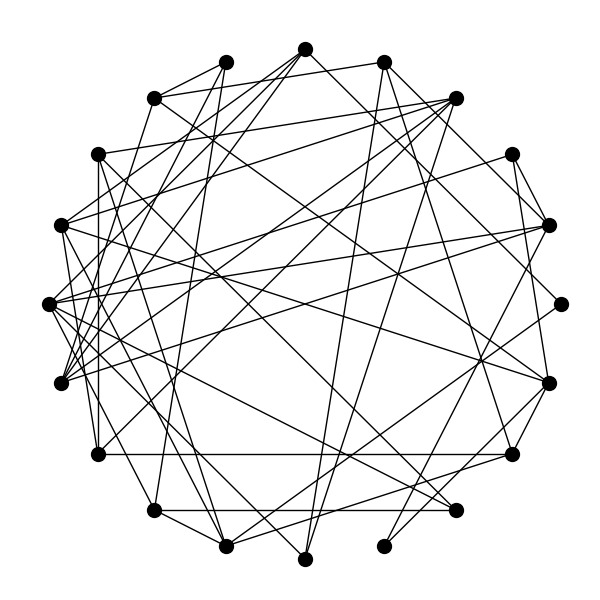}}
    \end{tabular}
\caption{\textbf{Network graphs.} Small graphs of size $N=20$ showing the different network structures: \textbf{(a)} Fully-Connected graph, \textbf{(b)} Scale-Free graph with initial size $m_0=5$, \textbf{(c)} Regular graph with node degree $k=4$, \textbf{(d)} Small-World graph with initial node degree $k=4$ and rewiring probability $p=0.2$ and \textbf{(c)} Random graph with initial node degree $k=4$. See text for more details.}
\label{fig_graphs}
\end{figure}
\item \textbf{Scale-Free networks.} They are networks whose degree distribution follows a power-law:
\begin{equation}
    P(k) \propto k^{-\gamma}.
\end{equation}
with $k$ the node degrees, $\gamma$ a real constant. The Barabási-Albert model gives a procedure for the construction of scale-free networks \cite{barabasi}, by starting with a small Fully-Connected network of $m_0$ nodes then adding one by one the $N-m_0$ remaining nodes, connecting them to the $m$ already present nodes with probability 
\begin{equation}
    p_v = \frac{k_v}{\sum\limits_w k_w},
\end{equation}
$w \in \llbracket 0:m-1\rrbracket$. Scale-free networks exhibit nodes with degrees that are several standard deviations away from the average degree of the network. These highly connected nodes are called \textit{hubs}. Note that, however, for smaller networks with size $N<100$, the scale-free property might not be properly observable.
\end{itemize}
Fig.~\ref{fig_graphs} provides a visual representation of the above mentioned graphs. Note that for visualization purposes we here show only a small fraction of the actual networks that we use later in our simulations where the number of nodes is set be $N=500$.

\subsection{The Kuramoto model}

We use of the Kuramoto model to study the neural activity of the coupled system. To this end we consider a population of $N$ phase oscillators \cite{Kura_1984}:
\begin{equation} \label{eq:KuraModel}
    \dot{\theta_i} = \omega_i + F\delta_{i}\sin(\Omega t + \theta_i) + \frac{K}{k_i}\sum_j A_{ij}\sin(\theta_j-\theta_i),
\end{equation}
where $\theta_i$ denotes the phase of the $i-$th oscillator, $\omega_i$ its respective frequency (Hz) drawn from a Lorentz probability distribution $g(\omega)$ of scale parameter $\gamma=0.5$, centered in $x_0=1$. $A$ is the binary adjacency matrix coupling the oscillators, $k_i$ the degree of oscillator $i$ 
and $K$ is the global coupling constant. We apply external stimulus in a subset of the oscillators with fixed amplitude $F$ and frequency $\Omega$. The term $\delta_{i}$ is a binary function indicating this subset of nodes where the stimulation is applied in different realization in our simulations,
\begin{equation} \label{eq:KuraModel_stim}
    \delta_{i} = \left\lbrace
    \begin{array}{ccc}
     1 & \mbox{if node $i$ is in the stimulated subset}\\
     0 & \mbox{else.}\\
    \end{array}\right.
\end{equation}
We set the time-step at $0.01$s and we integrated the system with an Euler scheme (no noise is considered). 

The system's degree of synchrony is measured using its order parameter $r$ \cite{Kura_1984}:
\begin{equation}
    re^{i\psi} = \frac{1}{N} \sum \limits_{j=1}^N e^{i\theta_j},
\end{equation}
where $\Psi$ denotes the population's mean phase. The order parameter $r$ tends to $1$ for a perfectly synchronized population and to $0$ in the absence of synchronization respectively. Due to the presence of strong fluctuations, all $r$ time-series shown in this paper are determined using a moving average on $r$, on time windows of length $2$s sliding each $0.1$s. The final states of a population, $r_f$, are computed by averaging these moving-averaged $r$ time-series over a $15$s time-window where the system has reached its stable state. The system's degree of synchrony depends on the coupling strength $K$'s relative position to a critical coupling strength $K_c$, whose value depends on the network configuration (see e.g. \cite{Mirollo_2007,Chiba_2018}). Here, we set this value at $K=0.2$ so that all considered networks are desynchronized in the absence of any external stimulation.

\section{Results}

\noindent \textbf{Network modulation.} In order to adequately tune the stimulus' amplitude and frequency values such that they can lead the system into a synchronous state, we first perform a systematic analysis in the parameter space $(F,\Omega)$. Hence, we begin by applying external stimulus to all the nodes for each pair of parameters and measure the final order parameter $r_f$. Such a parameter map reveals the presence of the well-known Arnold tongues \cite{Pikovsky_book_2001}, namely regions in the plane $(F,\Omega)$ for which the system gets synchronized.
\begin{figure}[h] \centering
\includegraphics[width=\textwidth]{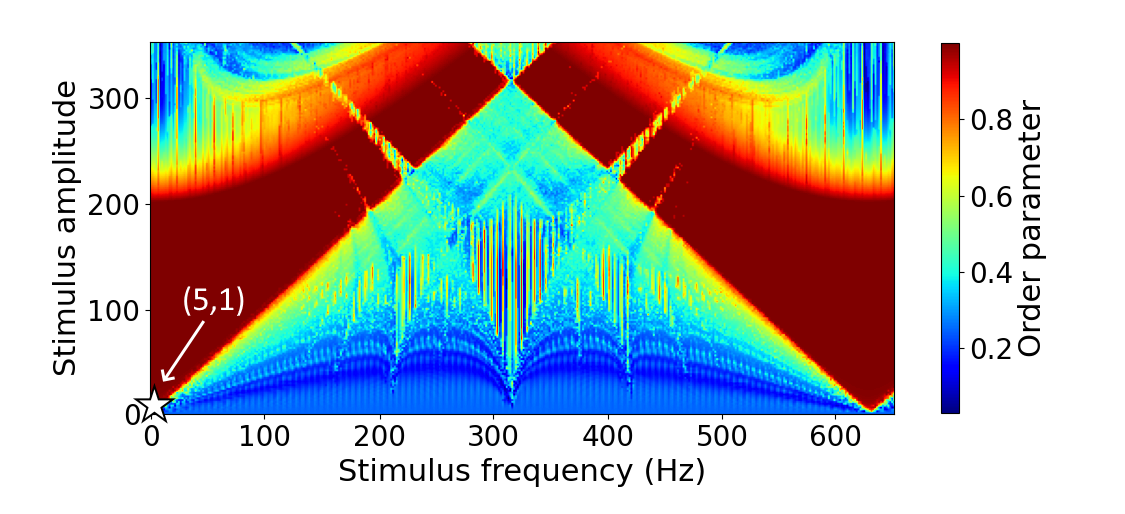}  
\caption{\textbf{Synchronization regions in the stimulation frequency-amplitude parameter space for a Regular network.} Final order parameter reached for a Regular network of size $N=20$, and degree $k=4$ where all nodes are stimulated, for different pairs of stimulus intensity ($F$) and frequency ($\Omega$) values in Eq.~(\ref{eq:KuraModel}). Each data point corresponds to a single simulation over $30$s, the final order parameter being averaged over the last $15$s. The color map shows large main synchronization regions, as well as small higher-order synchronization areas. The white star symbol at the bottom-left part of the figure indicates the chosen parameters $(F,\Omega)=(5,1)$ for the forthcoming simulations.}
\label{param_space}
\end{figure}

In Fig.~\ref{param_space}, we present the different synchronization regions (presence of several Arnold tongues) for a Regular network of a relatively small size $N=20$ and mean neighborhood $k=4$. For every other studied network, the maps depicts similar features with large synchronization regions at relatively small amplitudes of the external current, $F<200$. These tongues get thinner with higher values of $F$. Inside the main Arnold tongues, the oscillators are phase-locked at the forcing frequency $\Omega$ and $r_f\approx1$. Inside zones of weaker degree of synchrony $r_f<1$, some oscillators are phase-locked, while the oscillators of higher natural frequencies keep rotating independently. The white star symbol in the bottom-left part of the figure indicates the chosen parameters $(F,\Omega)=(5,1)$ for our forthcoming simulations, resulting in a partial phase-locking of the network. Note that we have performed similar analysis with larger sizes but smaller parameter grid size and the overall picture turns out to be consistent. We have also prepared similar plots for all considered network configurations (figures not shown here). \\

\noindent \textbf{Simulation protocol.} We set the values $F=5$, $\Omega=1$Hz for the stimulus intensity and frequency in Eq.~(\ref{eq:KuraModel}), so that the network is weakly entrained without being completely phase-locked. We then measure the degree of synchronization (with the order parameter) in different networks described in Sect.~\ref{sect:graphs}. The system starts evolving for 4s without any external input before we start applying the stimulation to a subset of nodes until 30s. More precisely, we stimulate different fractions of nodes, i.e., $25\%$, $50\%$ and $75\%$ in each given network. These nodes can be either chosen randomly, or depending on particular connectivity properties (as described in Sect.~\ref{sect:conn}). For the latter case, we first sort the nodes according to their connectivity relative measurements (from higher to lower), i.e. the eigenvector centrality, average shortest path length and clustering coefficient. The resulting time series are smoothed with a moving average, and their $r_f$ is averaged over the last $15$s. In order to obtain a statistically relevant value of the final value of the order parameter $r_f$, we performed $20$ simulations for each different network-setup (randomizing the initialization/generation of the networks, the natural frequencies and the initial conditions for each simulation). 

\begin{figure}[h] \centering
\begin{tabular}{lll}
\subfigure[Eigenvector centrality]{\label{sub1} \includegraphics[scale=0.19]{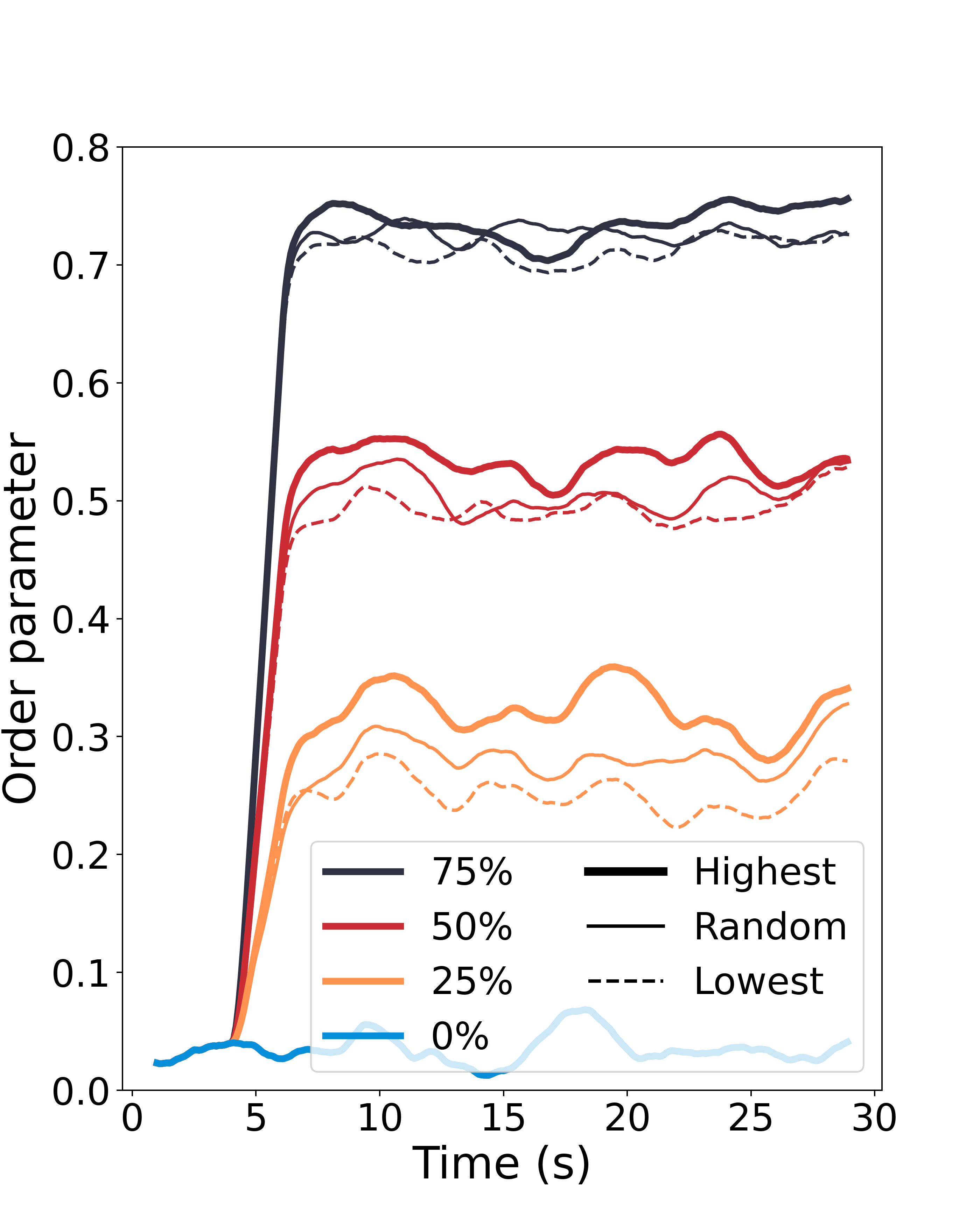}} &
\subfigure[Shortest path length]{\label{sub2} \includegraphics[scale=0.19]{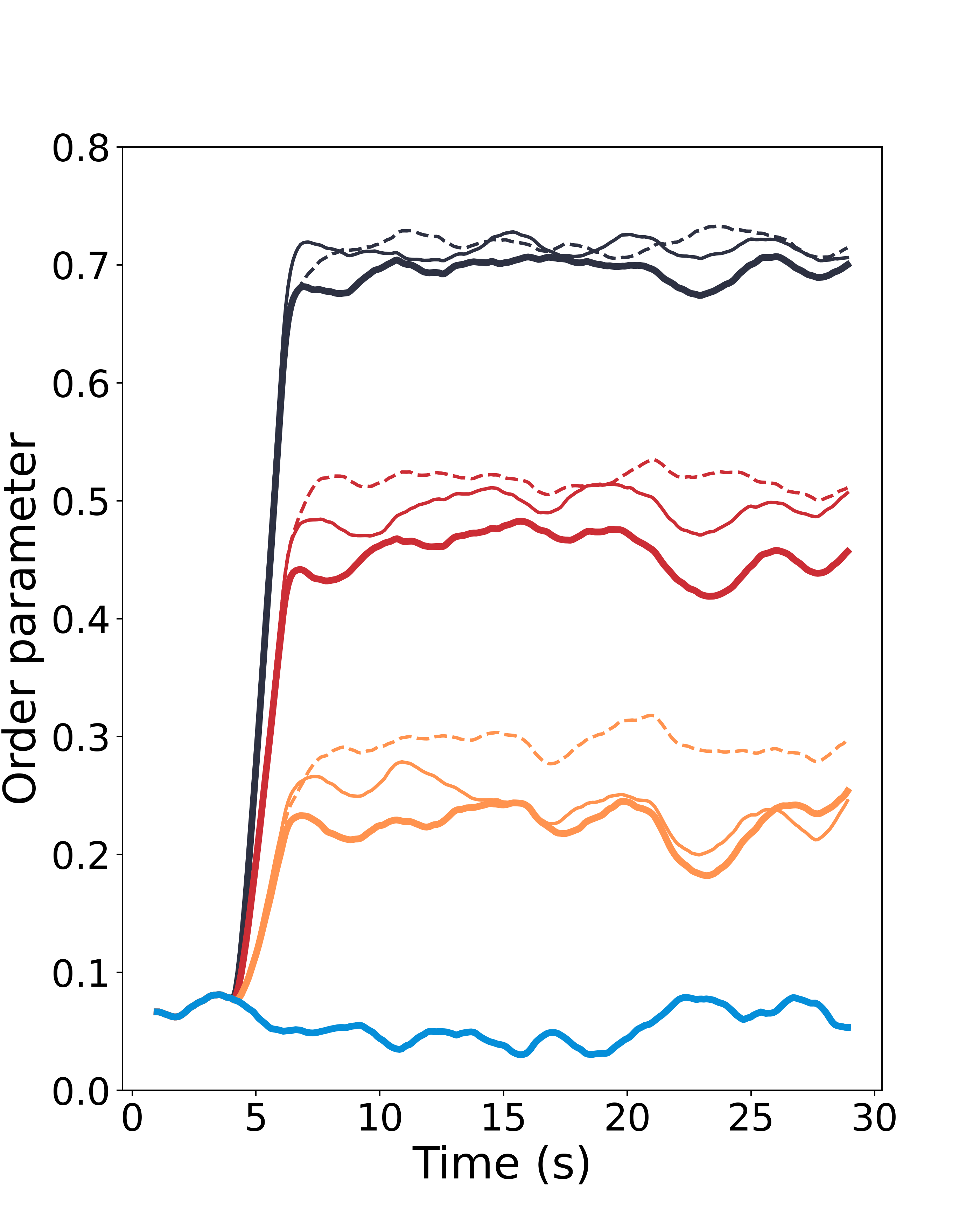}} &
\subfigure[Clustering coefficient]{\label{sub2} \includegraphics[scale=0.19]{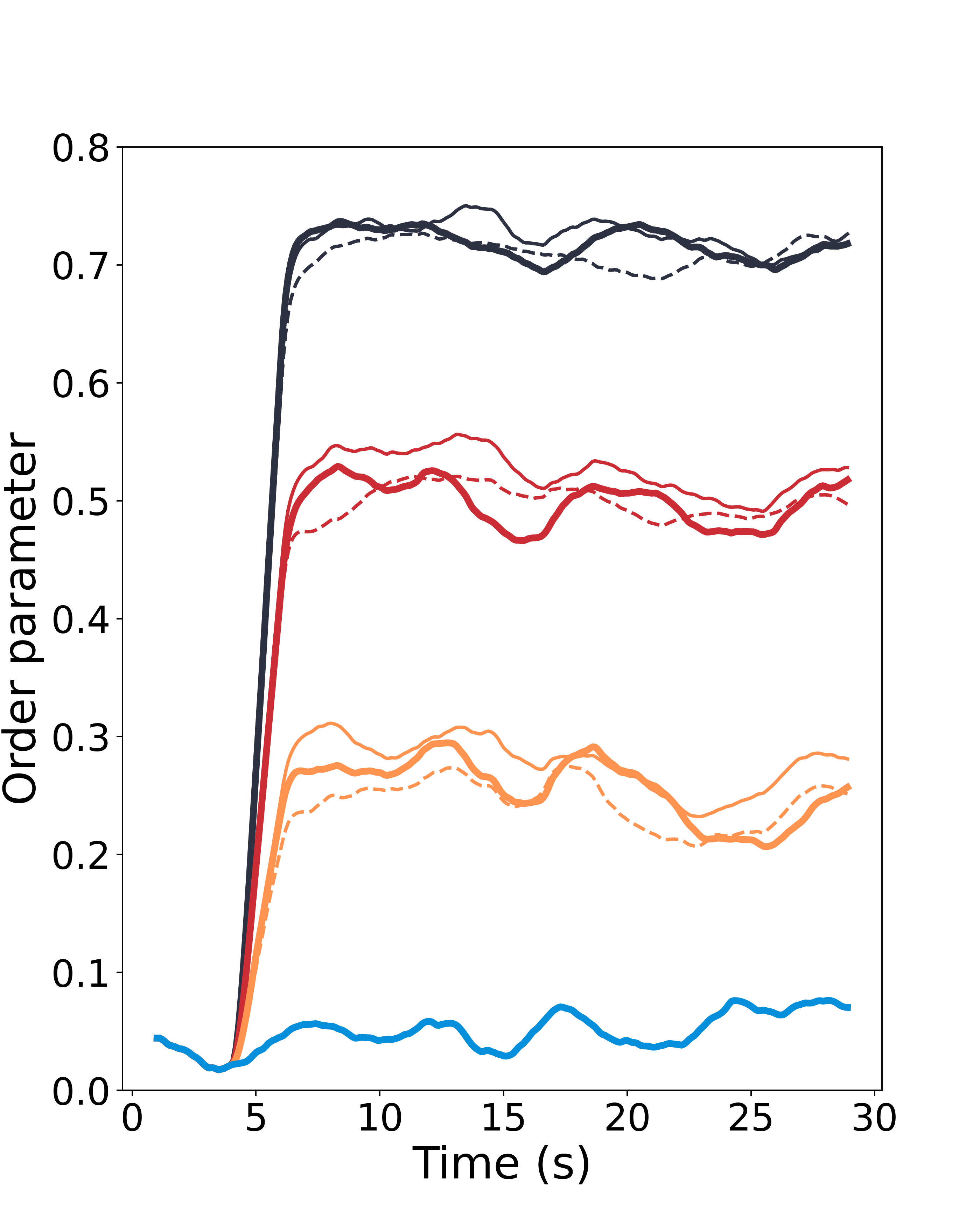}}
\end{tabular}
\caption{\textbf{Representative $r$ time-series for different stimulation setups of Scale-Free networks.} The order parameter as a function of time, for $N=500$ oscillators, when stimulating 0\% (blue), 25\% (orange), 50\% (red) and 75\% (black) of the nodes. The stimulated nodes are chosen randomly, then based on their \textbf{(a) }eigenvector centrality \textbf{(b) } average shortest path length and \textbf{(c) }clustering coefficient values. Bold solid lines (resp. dashed lines) correspond to the stimulation of nodes with the highest (resp. lowest) values, while thin solid lines correspond to the stimulation of randomly chosen nodes.}
\label{time_series}
\end{figure}

In Fig.\ref{time_series}, we show representative time-series for various stimulation setups for Scale-Free networks of size $N=500$ and initial size $m_0=5$. Each panel corresponds to a different initialization, from which the order parameter evolution is computed depending on the amount of stimulated nodes and the way they are selected. Thin solid lines correspond to randomly selected nodes. In that case, a first subset containing 25$\%$ of the nodes is created. Then for the stimulation of larger in size subsets, another 25$\%$ of nodes is successively added, so that larger subsets of random nodes always contains the smaller one. The bold solid lines (resp. dashed lines) show the time-series when the stimulated nodes are the ones with the highest (resp. lowest) eigenvector centrality (panel (a)), average shortest path length (panel (b)) and clustering coefficient (panel (c)). Note that higher values of the connectivity measurement do not necessarily lead to stronger synchrony. In particular, lower values for the nodes' average shortest path length depicts shorter connections to the rest of the network, and therefore a more efficient synchronization.\\

\begin{figure}[h!] \centering
\begin{tabular}{ll}
\subfigure[Small-World network]{\label{sub2} \includegraphics[scale=0.25]{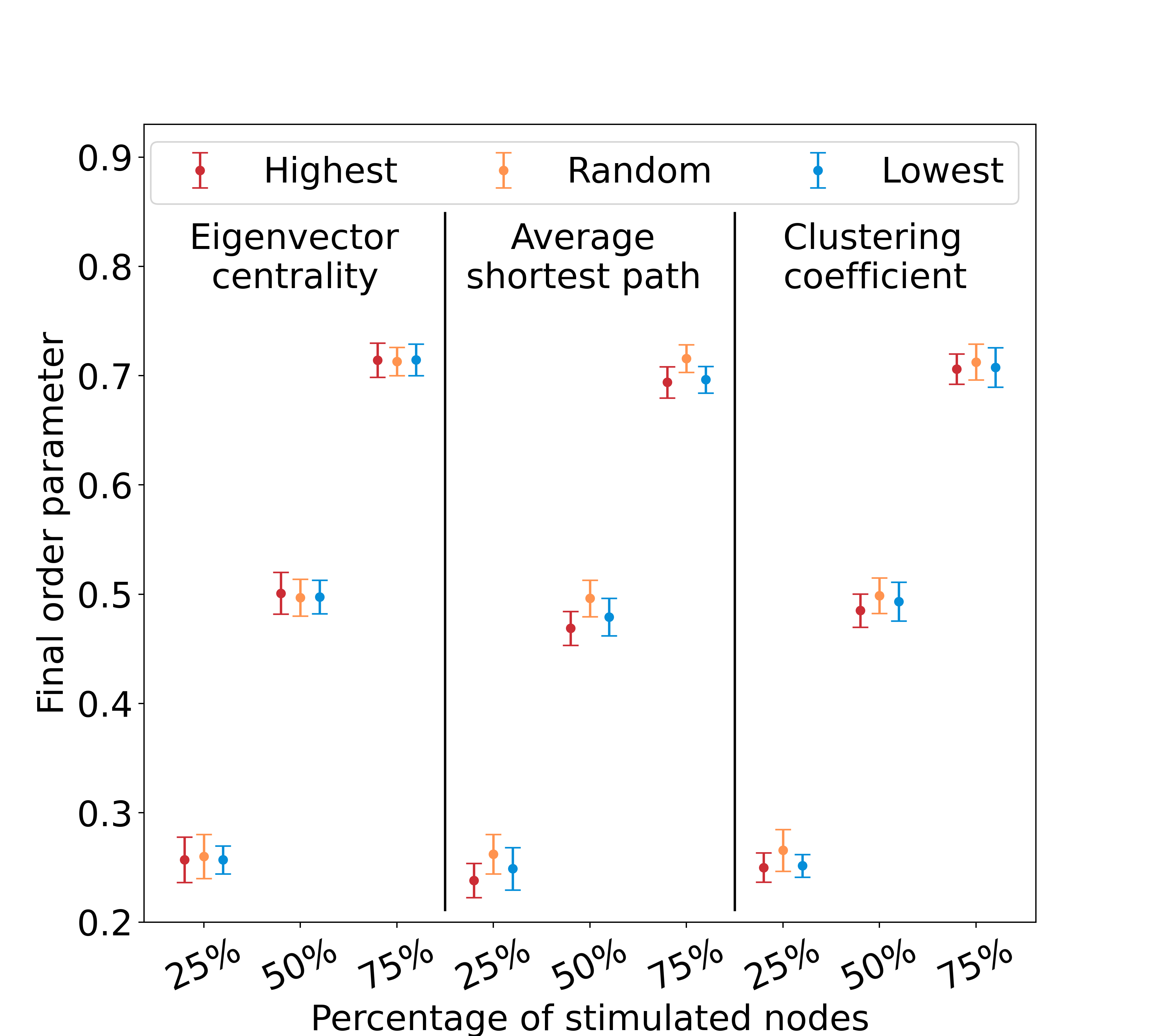}}\hspace{-0.5cm}
\subfigure[Random network]{\label{sub3} \includegraphics[scale=0.25]{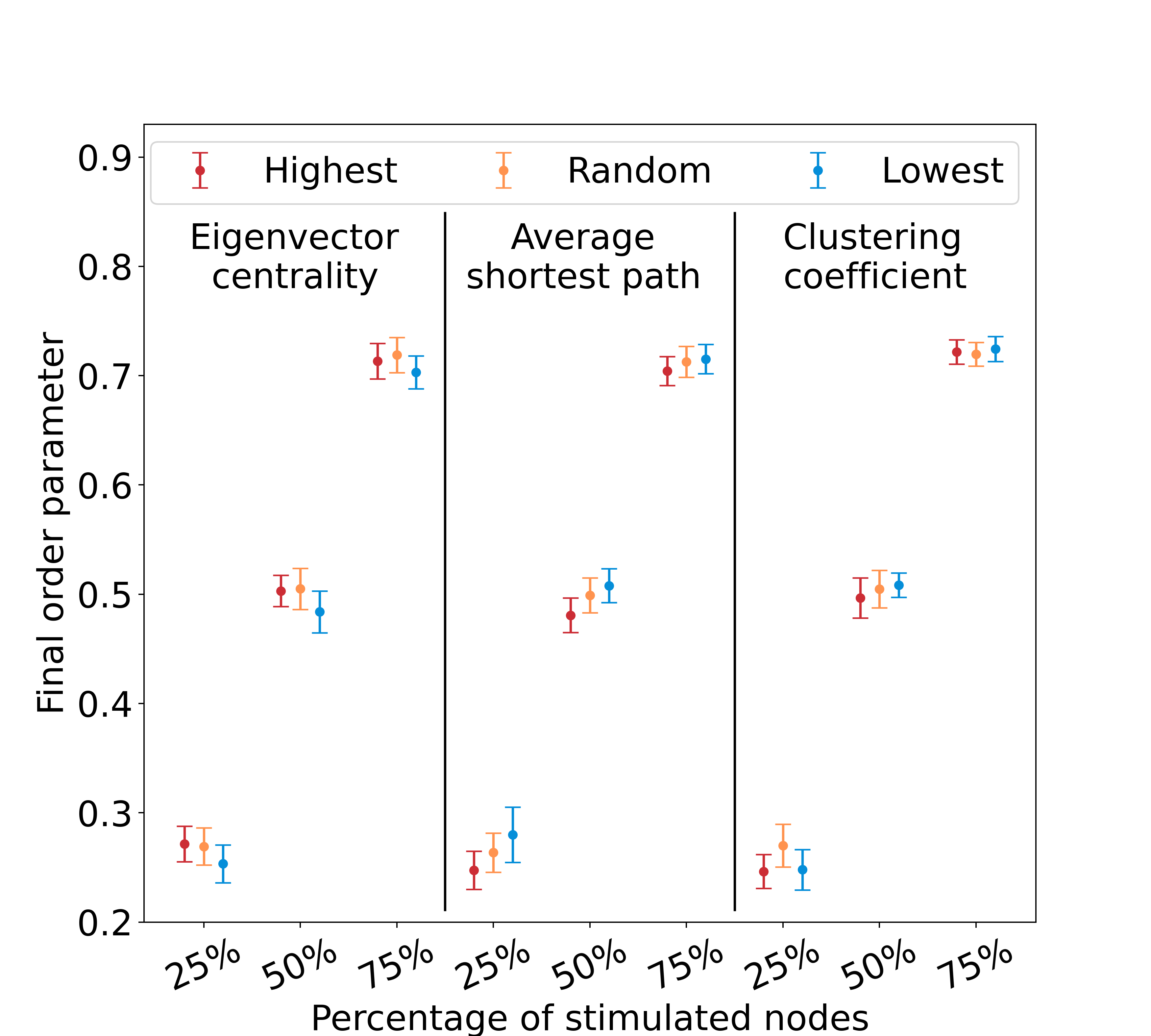}}
\end{tabular}
\subfigure[Scale-Free network]{\label{sub1} \includegraphics[scale=0.25]{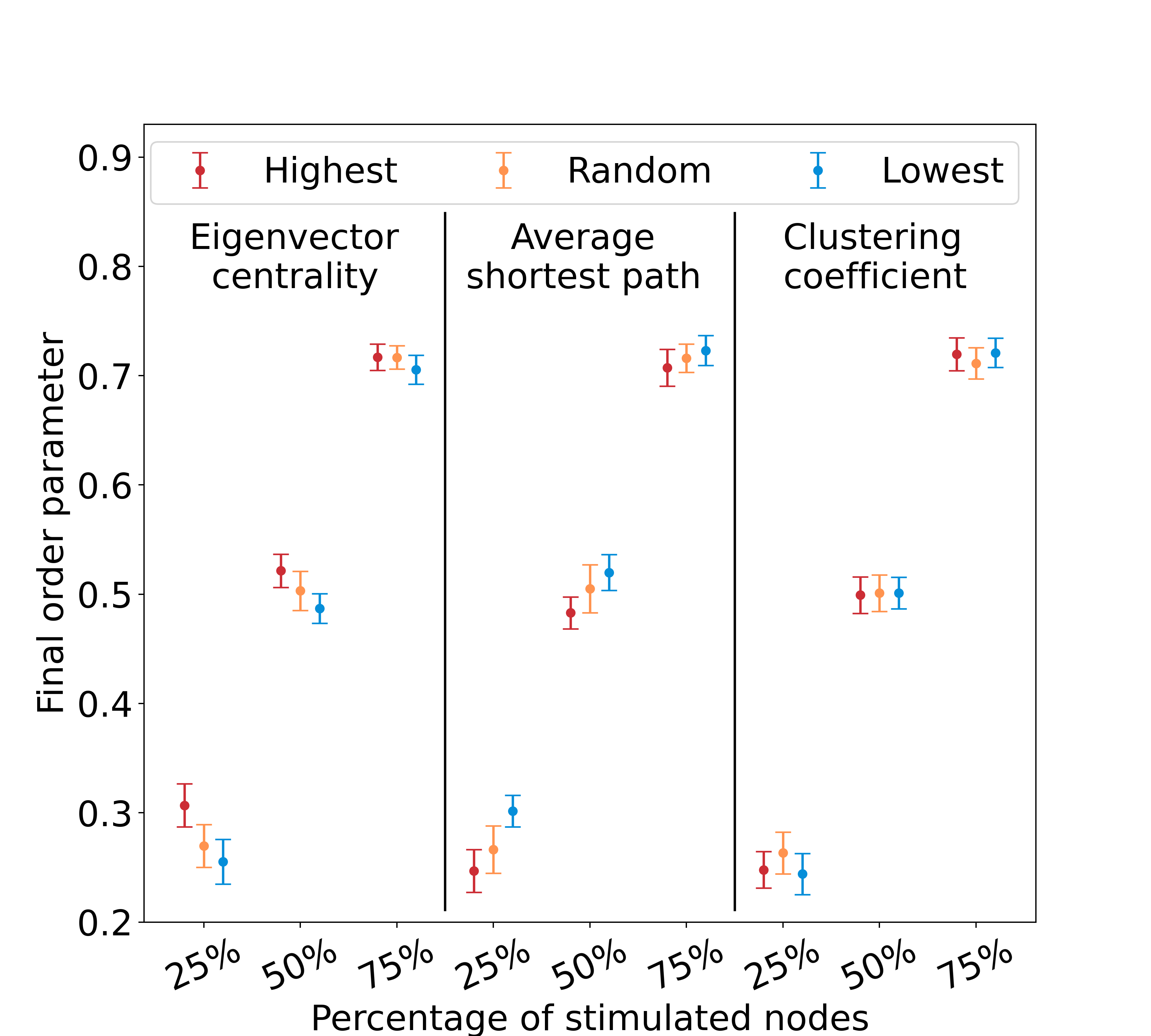}}
\caption{\textbf{Synchronization efficiency comparison for different types of graph networks.} Final order parameter obtained, for \textbf{(a) } Small-World networks \textbf{(b) }Random networks \textbf{(c) }Scale-Free networks of size $N=500$. The final value of the order parameter $r_f$ is computed for different stimulation subset sizes, composed of randomly chosen nodes (middle-orange boxplots in the legends), nodes with the highest connectivity measurement (left-red boxplots) and nodes with the lowest connectivity measurement (right-blue boxplots). $r_f$ shown are an average over $20$ simulations. The analysis is performed with three different connectivity measurements: eigenvector centrality (left column in each panel), average shortest path length (central column in each panel) and clustering (right column in each panel).}
\label{final_r}
\end{figure}

\noindent \textbf{Optimization of global synchronization}
In Fig.~\ref{final_r}, we present the main finding of our work, namely a systematic comparison of the synchronization efficiency when applying identical stimulus in different types of graph networks. Each panel is split into 3 columns showing the statistical summary for the ensembles of different realizations and choosing to stimulate different subsets of nodes, i.e. randomly (middle-orange boxplots in the legends) or with highest (upper-red boxplots in legends) or lower (lower-blue boxplots in the legends) connectivity  measurement. Panel (a) refers to a Small-World network of size $N=500$, initial degree $k=4$ and rewiring probability $p=0.2$, (b) to a Random network of size $N=500$, initial degree $k=4$ and rewiring probability $p=1$ and (c) to a Scale-Free network of size $N=500$ and initial size $m_0=5$ respectively. Note this analysis is not performed on Regular and Fully-Connected networks, since all nodes of such networks have identical connectivity properties and does not allow any connectivity-based selection.

In Scale-Free networks, the global order parameter reaches higher values when the stimulus is applied to the nodes with higher eigenvector centrality or lower average shortest path length. In such networks, a small fraction of the nodes have significantly higher connectivity, and stimulating preferentially these nodes enables strong synchronization compared to stimulating random nodes.

In Random networks, selecting stimulated nodes according to their lowest average shortest path length instead of randomly enhances synchronization. However, there is no benefit in choosing more central nodes. Indeed, the high degree of randomness (induced by a rewiring probability $p=1$, see Sect.~\ref{sect:graphs})  in these networks' connections causes disparity in the nodes' average shortest path length, without creating any node of way higher centrality. In Small-World networks, the connections are distributed in a more homogeneous way, and hence the node selection as no substantial impact on the system's final synchrony.

For all three aforementioned networks and all stimulation subset sizes, the selection of the nodes according to their clustering coefficient does not show any advantage over a simple random choice. Note that, we also checked for a potential overlap of clustering coefficient values between the randomly chosen nodes and those chosen with higher respective values. It turns out that they differ significantly. Our working hypothesis is that, by picking nodes with higher clustering coefficient, we induce higher synchrony locally (member nodes of the sub-network) while we disfavour other nodes outside of it. However, by choosing nodes randomly, we end up stimulating simultaneously nodes within and outside clusters eventually providing a better global synchronization. Finally, for all networks and stimulation subset selection methods, although they overall achieve higher degree of synchrony, larger stimulated subsets containing 75\% of the population do not allow to observe any clear advantage in particular selection of the nodes, since all three possible subsets largely overlap.

\section{Summary and Discussion}

In this study, we investigated the impact of structure and connectivity properties in a modulated network. We sought out to identify efficient ways to synchronize a population of Kuramoto phase oscillators using nodes' stimulation with fixed small amplitude and frequency. To this end, we first performed a parameter sweep exploration for stimulus amplitude and frequency parameters to identify settings that allow the system to synchronize. Then, we computed the evolution of characteristic networks of Kuramoto oscillators, where external stimulation is applied to different subpopulations with identical fixed low amplitude and frequency. In order to measure the system's synchrony of each network-type and stimulation configuration, we calculated the global order parameter for ensembles of different random system initializations.

We showed that by choosing this subpopulations based on their respective network connectivity properties (i.e. high eigenvector centrality and lower short-path length), we were able to enhance the networks' global degree of synchronization in comparison to the one achieved by randomly choosing them. This is not the case when using the clustering coefficient as a selection criterion. However, this direction  deserves further investigation, e.g. detect different network communities, measure their local synchronization and study its impact on the global synchrony when such nodes are chosen to be stimulated.

From a neuroscience point of view Scale-Free networks play an important role in the structure and function of mammal brains, see for example \cite{Ribeiro_etal_2021} (a study on the scale-free dynamics and the emergence of collective organisation occurs in rodents) or \cite{Grosu_etal_2022} (investigating the fractal structure of the human brain and its dynamics).
Furthermore, in Alzheimer patients' brain, the functional connectivity structure is found to exhibit properties similar to Random network graphs (see e.g. \cite{Stam_2009,Kang_2014} and references therein). Thus, understanding how to optimally synchronize systems with similar network structures can improve the overall expected performance of a given external simulation protocol.

\section*{Acknowledgements} JC was supported by the LABEX MME DII (ANR-16-IDEX-0008) PhD grant and CY cognition grant of the CY Cergy-Paris University. TM  was supported by the Labex MME DII (ANR-11-LBX-410 0023-01) French national funding program. MQ was supported by a CNRS financing for a research semester in the IPAL Lab in Singapore. We also acknowledge the use of the Osaka compute-cluster resources of CY Cergy-Paris University.

\section*{Code and data availability} All data generated or analysed during this study are included in this manuscript. Our code is written in \textit{Python} and we used the \textit{NetworkX} python package. Our code is available upon request.


\end{document}